\documentclass[12pt]{article}

\usepackage{latexsym}
\usepackage{amssymb}
\usepackage{amsbsy}
\usepackage{amsmath}
\usepackage{cite}
\usepackage{epsfig}
\usepackage{graphics,color}
\usepackage{a4}
\numberwithin{equation}{section}

\allowdisplaybreaks

\newcommand{\be}{\begin{equation}}
\newcommand{\ee}{\end{equation}}

\newcommand{\bea}{\begin{eqnarray}}
\newcommand{\eea}{\end{eqnarray}}
\newcommand{\bean}{\begin{eqnarray*}}
\newcommand{\eean}{\end{eqnarray*}}

\newcommand{\om} {\omega}

\begin{document}

\title{\bf\large
Melting of $P$ wave bottomonium states in the quark-gluon plasma from lattice NRQCD
}

\author{G.~Aarts${}^a$, C.~Allton${}^a$, S.~Kim${}^{b}$,  M.~P.~Lombardo${}^{c}$,  \\
S.~M.~Ryan${}^d$ and  J.-I.~Skullerud${}^e$ \\
\mbox{} \\
\mbox{} \\
{${}^a$\em\normalsize Department of Physics, Swansea University, Swansea, United Kingdom} \\
{${}^b$\em\normalsize Department of Physics, Sejong University, Seoul 143-747, Korea} \\
{${}^c$\em\normalsize INFN-Laboratori Nazionali di Frascati, I-00044, Frascati (RM) Italy} \\
{${}^d$\em\normalsize School of Mathematics, Trinity College, Dublin 2, Ireland} \\
{${}^e$\em\normalsize Department of Mathematical Physics, National University of Ireland Maynooth} \\ 
{\em\normalsize Maynooth, County Kildare, Ireland}
}

\date{October 21, 2013}

\maketitle

\begin{abstract}
We study the fate of $P$ wave bottomonium states in the quark-gluon plasma, using a spectral function analysis of euclidean lattice correlators. The correlators are obtained from lattice QCD simulations with two light quark flavours on highly anisotropic lattices, treating the bottom quark nonrelativistically. We find clear indications of melting immediately after the deconfinement transition.

\end{abstract}

%11.10.Wx, % Finite-temperature field theory
%12.38.Gc, % Lattice QCD calculations
%14.40.Pq % Heavy quarkonia

\maketitle

%%%%%%%%%%%%%%%%%%%%%%%%%%%%%%%%%%%%%%%%%%%%%%%%%%%

\newpage
 
\section{Introduction}
 \label{sec:intro}
 
 Ever since the suggestion that charmonium suppression may provide a signature for the formation of the quark-gluon plasma (QGP) in relativistic heavy-ion collisions  \cite{Matsui:1986dk}, quarkonia (heavy quark--antiquark bound states) immersed in the QGP have been studied intensely.
Due to the large energies available at the Large Hadron Collider, not only charmonium but also 
bottomonium can now be investigated experimentally \cite{Chatrchyan:2011pe,:2012fr,Dahms:2013eaa,ALICE:2013kwa} 
and this has led to an increasing theoretical interest in understanding the bottomonium system and its phenomenology at finite temperature, see e.g.\ the recent reviews \cite{Mocsy:2013syh,Rothkopf:2013ria,Ghiglieri:2013iya} and references therein, as well as Refs.\ \cite{Agotiya:2009aq,Sharma:2012dy,
Suzuki:2012ze,Dominguez:2013fca}.

In a series of papers \cite{Aarts:2010ek,Aarts:2011sm,Aarts:2012ka} we have studied bottomonium at nonzero temperature using simulations of lattice QCD, in which the heavy $b$ quarks propagate nonrelativistically through a quark-gluon plasma 
with $N_f=2$ flavours. We use a highly anisotropic lattice formulation, with a total of seven different temperatures up to $2.1T_c$. In Refs.\ \cite{Aarts:2011sm,Aarts:2012ka} we focused on the $S$ waves (in the $\Upsilon$ and $\eta_b$ channels) at zero and nonzero momentum, and found that the ground states  survive up to the highest temperature available, while excited states appear to dissolve close to $T_c$.
These results are consistent with experimental findings at the LHC  \cite{Chatrchyan:2011pe,:2012fr,Dahms:2013eaa,ALICE:2013kwa}. In Ref.\ \cite{Aarts:2010ek} first results for $P$ waves were presented (in the $\chi_b$ channels). In the QGP phase we observed clear indications for nonexponential decay of the euclidean lattice correlators, which was interpreted as the melting of the ground (and excited) states.

Our aim for this paper is to complete the analysis of the $P$ wave states. In Ref.~\cite{Aarts:2010ek} we studied correlators (and no spectral functions) at only four temperatures with limited statistics: here we extend this to the full ensemble also used in Refs.\ \cite{Aarts:2011sm,Aarts:2012ka} and provide the spectral functions, obtained with the help of the Maximum Entropy Method (MEM). 
The paper is organised as follows. The correlators and spectral functions are presented in Secs.\ \ref{sec:corr} and \ref{sec:rho} respectively. Sec.~\ref{sec:sys} contains a discussion of systematics in the MEM analysis. Finally, a summary is given in Sec.~\ref{sec:sum}.

 \section{Correlators}
 \label{sec:corr} 
 
 We use the formulation and lattice ensembles discussed in Ref.\ \cite{Aarts:2011sm} and refer to that paper for further details. Here we mention that the applicability of nonrelativistic QCD (NRQCD) 
\cite{Caswell:1985ui}  on the lattice \cite{Davies:1994mp} 
to treat the $b$ quark at finite temperature is motivated by the hierarchy of effective field theories formulated in Refs.\ \cite{Laine:2006ns,Burnier:2007qm,Brambilla:2010vq,Brambilla:2011sg}. 
We also emphasise that the use of NRQCD enhances the signature for quarkonium melting/survival, as it avoids several problems which have complicated the study of relativistic quarks in thermal equilibrium  \cite{Umeda:2007hy,Petreczky:2008px}. In particular,  constant contributions associated with transport and susceptibilities \cite{Aarts:2002cc} are not present \cite{Aarts:2010ek}. Moreover, the entire euclidean time interval can be used, due to the absence of backward moving states. 
   For completeness, some lattice parameters are provided in Table~\ref{tab:lattice}.

\begin{table}[t]
\begin{center}
\vspace*{0.2cm}
\begin{tabular}{| l | rrrrrrr | }
\hline
$N_\tau$ 		& 80 & 32 & 28 & 24 & 20 & 18 & 16 \\
$T$(MeV) 	& 90 & 230 & 263 & 306 & 368 & 408 & 458 \\
$T/T_c$ 		& 0.42 & 1.05 & 1.20 & 1.40 & 1.68 & 1.86 & 2.09 \\ 
$N_{\rm cfg}$   & 250 & 1000 & 1000 & 500 & 1000 & 1000 & 1000 \\
\hline
\end{tabular}
\vspace*{0.2cm}
\caption{Two-flavour lattice details: the lattice size is $N_s^3\times
  N_\tau$ with $N_s=12$, lattice spacing $a_s\simeq 0.162$ fm, $a_\tau^{-1} =7.35(3)$ GeV, and anisotropy $a_s/a_\tau=6$ \cite{Aarts:2011sm}.  
}
\label{tab:lattice}
\end{center}
\end{table}

\begin{figure}[t]
\begin{center}
\epsfig{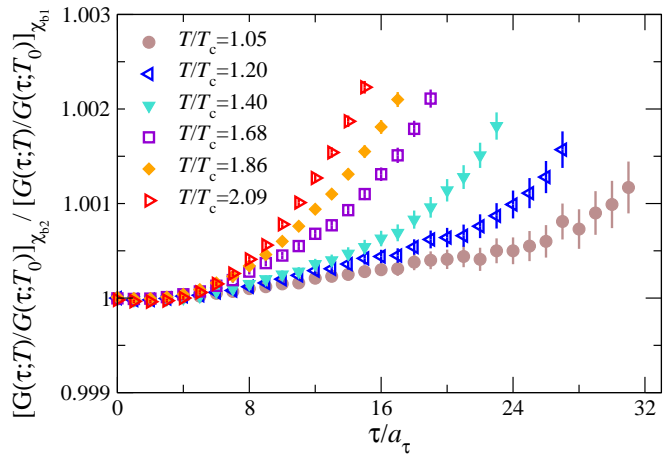}
\end{center}
 \caption{Ratio  $[G(\tau;T)/G(\tau;T_0)]_{\chi_{b2}}$  of the correlator in the $\chi_{b2}$ channel at a given temperature $T$ with the one at the lowest temperature, $T_0=0.42T_c$, divided by the same quantity in the  $\chi_{b1}$ channel,  $[G(\tau;T)/G(\tau;T_0)]_{\chi_{b1}}$.
}
 \label{fig:0}
\end{figure}

We start with a discussion of the correlators $G(\tau)$. All correlators are at zero momentum and we use point sources throughout. We have analysed $P$ wave correlators in the 
  $^3P_0$(scalar, $\chi_{b0}$), 
  $^3P_1$(axial-vector, $\chi_{b1}$), 
  $^3P_2$(tensor, $\chi_{b2}$),
and $^1P_1$($h_b$)  channels. 
We found that the correlators in the different channels behave in a very similar way. To illustrate this, we show in Fig.\ \ref{fig:0} double ratios, namely the ratios of the correlator $G(\tau;T)$ in the $\chi_{b2}$ channel at a temperature $T$, normalised by the correlator at the lowest temperature, $T/T_c=0.42$, divided by the same quantity in the $\chi_{b1}$ channel. 
As can be seen, the difference, although statistically significant, is at most a few per mille at the largest euclidean times.
Hence from now on we present results in the $\chi_{b1}$ channel only.

\begin{figure}[t]
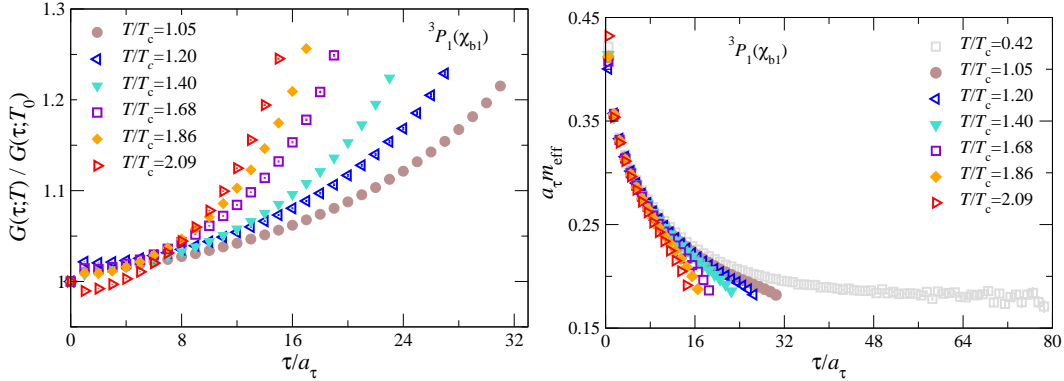

\begin{center}
\epsfig{figure=chi_b1_C80-GA.eps,width=0.48\textwidth}
\epsfig{figure=chi_b1_all-T_eff-mass-GA.eps,width=0.48\textwidth}
\end{center}
 \caption{Left: Ratio of the correlator $G(\tau;T)$ at a given temperature $T$ with the one at the lowest temperature, $T_0=0.42T_c$,  in the $\chi_{b1}$ channel.
 Right: euclidean-time dependence of the effective mass, for different temperatures.
}
 \label{fig:1}
\end{figure}

In Fig.\ \ref{fig:1} (left) we show single ratios: the correlation functions at a given temperature $T$, normalized by the correlator at the lowest temperature, $T/T_c=0.42$, as a function of the euclidean time. We observe a substantial temperature dependence in the whole temperature range. The maximal deviation, at the largest $\tau$ value for each temperature, is around 20-25\%. This should be contrasted with the situation for $S$ waves, where we found a maximal deviation of less than 3\%, see Fig.\ 1 of Ref.\ \cite{Aarts:2011sm}. This is the first indication of strong temperature effects for $P$ waves. We reiterate that this temperature dependence is not due to changes in the susceptibility or zero mode \cite{Petreczky:2008px}, since this contribution is absent in NRQCD.
 
In Fig.\ \ref{fig:1} (right) we show the effective masses, 
\be
a_\tau m_{\rm eff}(\tau) = - \log[G(\tau)/G(\tau-a_\tau)],
\ee 
as a function of $\tau$. When the correlator takes the form of a sum of exponentials, the ground state will show up as a plateau at large euclidean times, provided that it is well separated from the excited states. This is indeed the case at the lowest temperature and leads to the zero-temperature spectrum discussed in Refs.\  \cite{Aarts:2010ek,Aarts:2011sm}. Above $T_c$, we observe that the effective masses no longer follow the trend given by the correlator below $T_c$, but instead bend away from the low-temperature data.
This implies that the spectrum has changed drastically. If isolated bound states persist, the ground state has to be much lighter and excited states cannot be well separated. A more natural explanation is that there is no exponential decay and bound states have melted, immediately above $T_c$. This interpretation is supported by the spectral function analysis presented next.

\section{Spectral functions}
\label{sec:rho}

To obtain further insight into how $P$ wave states are modified in the QGP, we now turn to a spectral-function analysis, similar to that presented in Refs.\ \cite{Aarts:2011sm,Aarts:2012ka}; we refer to those papers for details. In NRQCD the spectral relation reads  \cite{Burnier:2007qm,Aarts:2010ek}
\be
\label{eq:spec}
G(\tau) = \int_{\om_{\rm min}}^{\omega_{\rm max}} \frac{d\omega}{2\pi}\, K(\tau,\om)\rho(\omega),
\quad\quad\quad\quad
K(\tau,\om)\ = e^{-\omega\tau},
\ee 
both at zero and nonzero temperature, unlike the case of relativistic quarks, where the kernel $K(\tau,\om)$ is temperature dependent. 
We invert Eq.\ (\ref{eq:spec}) with the help of MEM \cite{Asakawa:2000tr,Bryan}, using a constant default model, with limits $a_\tau\om_{\rm min}\sim 0$ and $a_\tau\om_{\rm max}\sim 2$  \cite{Aarts:2011sm}.
Since the heavy quark mass scale is integrated out, energies are determined up to an overall additive shift, which is fixed by comparing the spectrum at zero temperature with the actual physical spectrum. Specifically, we used the $\Upsilon(1S)$ mass. We therefore identify $a_\tau\om=0$ with $\om=8.57$ GeV \cite{Aarts:2011sm}.

\begin{figure}[h]
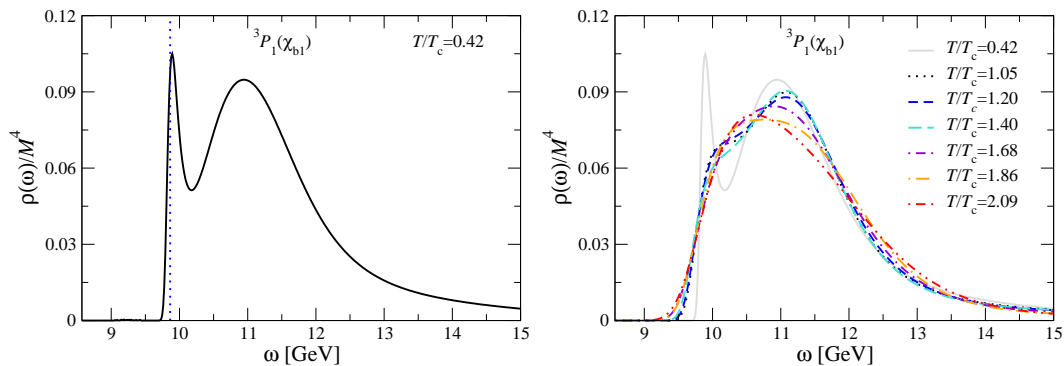

\begin{center}
\epsfig{figure=rho-chi_b1_Nt80_we-rescaled-GA.eps,width=0.48\textwidth}
\epsfig{figure=rho-chi_b1_all_woe-rescaled-GA.eps,width=0.48\textwidth}
\end{center}
 \caption{Spectral function in the $\chi_{b1}$ channel at the lowest temperature (left) and at all temperatures (right). The dotted line on the left indicates the position of the ground state  obtained with a standard exponential fit.
 }
 \label{fig:2}
\end{figure}

The result at the lowest temperature is given in Fig.\ \ref{fig:2} (left). The dotted vertical line indicates the mass of the lowest-energy state obtained with an exponential fit. We see from this that the narrow peak in the spectral function corresponds to the ground state. 
The second, wider structure is presumably a combination of excited states and lattice artefacts, see below. We note that we have not been able to extract the mass of the first excited state with an exponential fitting procedure.
The spectral functions for all temperatures are shown in Fig.\ \ref{fig:2} (right). We 
find no evidence of a ground state peak for any of the temperatures above $T_c$. This is consistent with the interpretation of the correlator study presented above and supports the conclusion that the $P$ wave bound states melt in the QGP.

\begin{figure}[t]
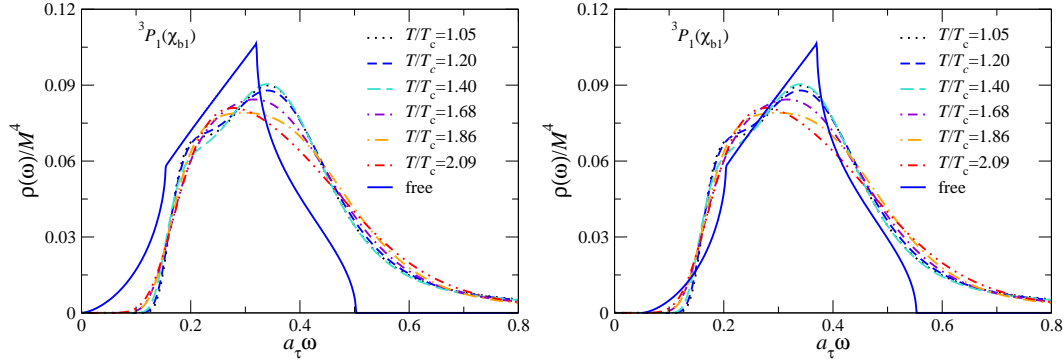

\begin{center}
\epsfig{figure=rho-chi_b1_all-aw-free1-GA.eps,width=0.48\textwidth}
\epsfig{figure=rho-chi_b1_all-aw-free2-GA.eps,width=0.48\textwidth}
\end{center}
 \caption{Comparison with the free lattice spectral function above $T_c$ with the threshold at $a_\tau\om=0$ (left) and slightly shifted (right).
 }
 \label{fig:3}
\end{figure}

In order to interpret the remaining structure, we compare it with the spectral function computed on the lattice, in the absence of interactions \cite{Aarts:2011sm}. 
In the continuum the free spectral function in the $P$ wave channel is given by \cite{Burnier:2007qm}
\be
\rho_P^{\rm free, cont}(\om) \sim \left(\omega-\om_0\right)^{3/2}\Theta(\om-\om_0),
\ee
where $\om_0$ is the two-quark threshold. Note that in NRQCD the free spectral function is independent of the temperature. On the lattice this simple expression is modified due to lattice artefacts: in particular,  since the lattice momenta take values in the first Brillouin zone only, it has support in a finite energy range. Moreover the edges of the Brillouin zone show up as cusps in the free lattice spectral function, see Appendix A of Ref.\ \cite{Aarts:2011sm} for details.

In Fig.\ \ref{fig:3} we show the free lattice spectral functions, together with the spectral functions above $T_c$. In NRQCD the area under the spectral function is given by the source at $\tau=0$ \cite{Aarts:2011sm}; we have adjusted the overall normalisation of $\rho_P^{\rm free, lat}(\om)$
to find approximate agreement.  In principle the threshold $\om_0$, where  $\rho_P^{\rm free, lat}(\om)$ starts to increase, is determined by twice the heavy quark mass. In the free calculation, this coincides with $a_\tau\om=0$, as in Fig.\ \ref{fig:3} (left), but in the interacting theory this value will depend on details of the lattice simulations and the heavy quark mass, and will in general be different. Therefore we are allowed to shift $\rho_P^{\rm free, lat}(\om)$ horizontally, as in Fig.\ \ref{fig:3} (right).
On the other hand, what is not adjustable is the width of the region where $\rho_P^{\rm free, lat}(\om)$ has support and the position of the cusps. Therefore these features can be sensibly compared between the interacting and the free theory: we observe that the structure of the full spectral functions from NRQCD simulations in the QGP is not too dissimilar to that of $\rho_P^{\rm free, lat}(\om)$, another indication of unbound, quasi-free $b$ quarks.
This lends further support to the conclusion drawn in Ref.\  \cite{Aarts:2010ek} from an analysis of the correlators, namely that
the system in the $P$ wave channels is approaching a system of noninteracting quarks.

\section{Systematics} 
\label{sec:sys}

To ensure the robustness of the results produced with MEM, it is necessary to investigate the dependence on the various input variables in the MEM analysis, such as the default model, the precision of the data and the time range used. A detailed study of systematic effects in the MEM analysis in the case of  $S$ wave channels can be found in our previous paper \cite{Aarts:2011sm}. We have repeated that analysis for the $P$ wave channels considered here and found that  the results are again robust against variation of most of the input variables. However, we have found one 
aspect that is specific for the $P$ wave channels, namely a very strong dependence on the choice of final time slice. In contrast, in the $S$ wave channels the results do not depend strongly on the final time slice.

\begin{figure}[h]
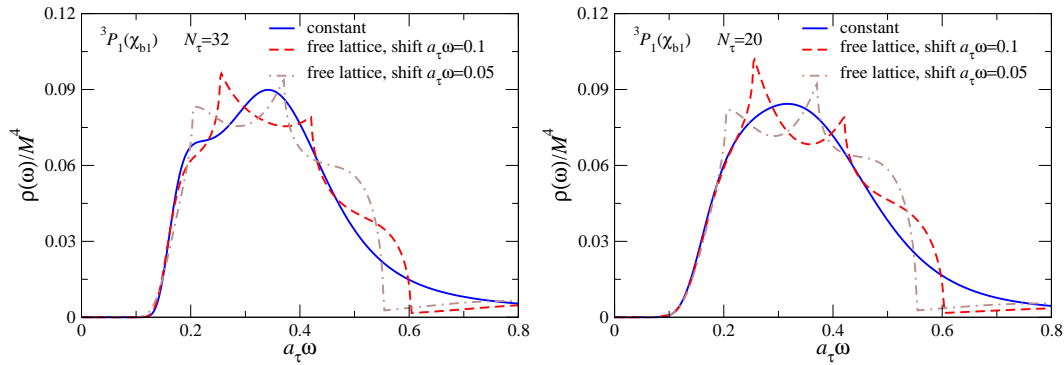

\begin{center}
\epsfig{figure=rho-chi_b1_free-recon-32-GA.eps,width=0.48\textwidth}
\epsfig{figure=rho-chi_b1_free-recon-20-GA.eps,width=0.48\textwidth}
\end{center}
 \caption{Reconstructed spectral functions using a constant default model and the free lattice default model, shifted by $a_\tau\om=0.05$ and 0.1, in the $\chi_{b1}$ channel, for $N_\tau=32$ (left) and 20 (right). 
 }
 \label{fig:def}
\end{figure}

We start with a discussion of the default model dependence. The default model $m(\om)$ enters in the parametrisation of the spectral function, 
\be
\rho(\om) = m(\om) \exp \sum_k c_ku_k(\om),
\ee
where $u_k(\om)$ are basis functions fixed by the kernel $K(\tau,\om)$ and the number of time slices, while  the coefficients $c_k$ are to be determined by the MEM analysis \cite{Asakawa:2000tr}. We find that the results are insensitive to the choice of default model, provided that it is a smooth function of $\om$. It is also possible to use as default model the free lattice spectral function, which has cusps and  support in a finite $\om$ interval only, as shown above. To use $\rho_P^{\rm free, lat}(\om)$ as default model, we add a small constant so that it is nonzero in the entire $\om$ range. We also shift it in the horizontal direction, such that the thresholds are located at $a_\tau\om=0.05$ and 0.1.
The resulting spectral functions are shown in Fig.\ \ref{fig:def}, together with the results from the constant default model, $m(\om)=m_0$. We observe that the nonanalytical behaviour of the free lattice default models is still present in the reconstructed spectral functions, since the basis functions $u_k(\om)$ are not capable of eliminating this. Shifting the default model simply results in a shift of these cusps.
From this we conclude that using a lattice free default model introduces a bias which cannot be overcome. We hence favour choosing smooth default models only: the results of the previous section are obtained with a constant default model. We also see that lattice artefacts potentially set in quite close to the threshold, which can be avoided by using lattices with a finer (spatial) lattice spacing: this will be necessary to 
clearly separate the physical region from the region dominated by lattice artefacts.
Nevertheless,  we observe that the position of the threshold agrees between the three default models. Moreover, as already noted above, the spectral functions reconstructed using a constant default model have support in the region marked by the free lattice spectral function. This result seems therefore clearly encoded in the euclidean correlator data.

\begin{figure}[t]
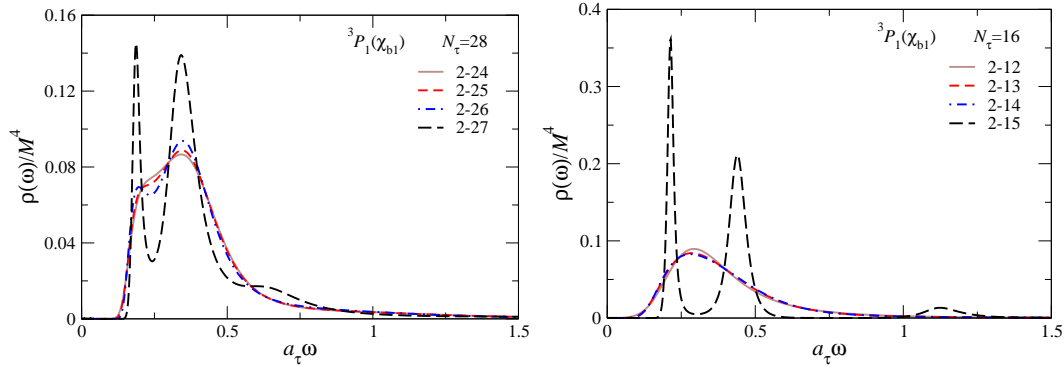

\begin{center}
\epsfig{figure=chi_b1_t2-t27_28_GA.eps,width=0.48\textwidth}
\epsfig{figure=chi_b1_t2-t15_16_GA.eps,width=0.48\textwidth}
\end{center}
 \caption{Dependence of the reconstructed spectral function on the final time slice used in MEM: $\tau/a_\tau\in [n_1,N_\tau-n_2]$, with $n_1=2$ and $n_2 = 4,3,2,1$, for $N_\tau=28$ (left) and 16 (right). The source is located at $n_1=0$.
 }
 \label{fig:t2}
\end{figure}

\begin{figure}[t]
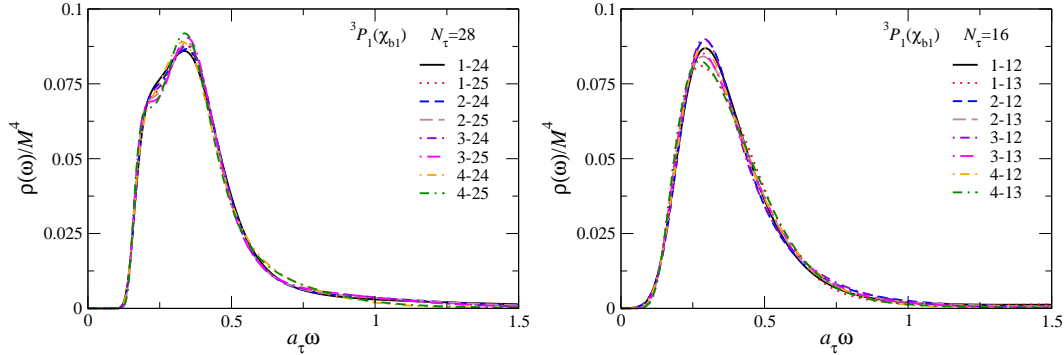

\begin{center}
\epsfig{figure=chi_b1_t1234-t2425_28_GA.eps,width=0.48\textwidth}
\epsfig{figure=chi_b1_t1234-t1213_16_GA.eps,width=0.48\textwidth}
\end{center}
 \caption{As in Fig.\ \ref{fig:t2}, with $n_1=1,2,3,4$ and $n_2=4,3$. 
 }
 \label{fig:t1}
\end{figure}

We now turn to the dependence on the time range. 
In the MEM analysis, one chooses the time interval, which should be varied in order to establish robustness of the output. 
We recall that in NRQCD the correlators are not periodic in time,\footnote{Nonsymmetric correlators also appear in the case of thermal Wilson loops, used to construct heavy-quark potentials \cite{Rothkopf:2011db,Bazavov:2013zha}.}
 so that in principle the entire time interval can be used, rather than only up to $\tau=1/2T$. Thermal effects appear since the $b$ quarks propagate through an ensemble of gluons and light quarks at finite temperature, which is implemented via the usual (anti)periodic thermal boundary conditions. Hence, even though boundary conditions do not appear explicitly in the propagation of the $b$ quarks, they are still present indirectly through the interaction with the QGP. 
To test the dependence on the time interval, we write the (discretised) time interval as
\be
\tau/a_\tau = n_1, \ldots, N_\tau-n_2,
\ee
where the source is located at $n_1=0$ and the maximal time is $N_\tau-1$, i.e.\ $n_2=1$ (and not $N_\tau/2$). In Fig.\ \ref{fig:t2} we present the dependence on the final time slice, varying $n_2=4,3,2,1$ at a fixed $n_1=2$, for $N_\tau=28$ (left) and 16 (right). Interestingly, we observe a clear double peak structure when the largest possible time is used, $n_2=1$. Moreover, the peaks get more pronounced as the temperature is increased. However, the double peak structure is only seen when $n_2=1$ and is not robust under variation of $n_2$. To illustrate this, we show in Fig.\ \ref{fig:t1} results with $n_2=3,4$ and varying $n_1=1,2,3,4$. In this case the output is robust and we recover the spectral functions discussed above. In particular, there is very little dependence on the initial time slice used.
While we not fully understand why the inclusion of the entire time interval has this effect, 
we conclude that the  behaviour found when $n_2=1$ is anomalous: it presents a thermal lattice artefact due to the periodicity of the gluonic fields through which the $b$ quarks propagate and  it should be discarded. 
For completeness we note that the spectral functions shown in Figs.\ \ref{fig:2} and \ref{fig:3} are obtained with $n_1=1$ and $n_2=3$.

\section{Summary}
\label{sec:sum}

We found strong indications that bottomonium states in the $P$ wave channels ($\chi_b$ and $h_b$) melt in the QGP, at a temperature close to $T_c$. This is in contrast to the situation in the $S$ wave channels ($\Upsilon$ and $\eta_b$) where we found previously that the ground states survive up to the highest temperature we consider, $T/T_c=2.09$. The conclusion for $P$ waves is based on observations of the correlators and the associated spectral functions, which we find to be consistent: an absence of a plateau in the effective masses in the first case and the absence of an isolated ground state peak in the second. Moreover, above $T_c$ the spectral functions are similar in shape to those found for free nonrelativistic quarks on the lattice.
As a technical remark, we note that the MEM results are robust under variation of the euclidean time interval used, provided that the final time slice is not included. 
We emphasise that our results have been obtained on ensembles with a single lattice spacing: systematic effects related to lattice spacing dependence have not yet been addressed.

As an outlook, we note that there now exist various alternatives to MEM  \cite{Rothkopf:2011ef,Burnier:2013nla} and it would be interesting to apply those techniques in order to gain further confidence in our conclusions, also on ensembles generated with different lattice actions \cite{seyong}.
Similarly,  it would be interesting to compare our results with potential model calculations. 
Finally, we are in the process of extending this work to ensembles with $N_f=2+1$ dynamical quarks, using lattices with an anisotropy of $a_s/a_\tau=3.5$ and a finer spatial lattice spacing. First results on these $N_f=2+1$ ensembles have been obtained for the electrical conductivity \cite{Amato:2013naa} and susceptibilities \cite{Giudice:2013fza}; the study of bottomonium is currently under investigation \cite{tim}.

\section*{Acknowledgments}

We thank Don Sinclair and Bugra Oktay for collaboration on related projects.
CA and GA thank Trinity College Dublin for hospitality during the course of this work. 
GA thanks TIFR Mumbai for its tranquility during the completion of the manuscript. 
This work was partly supported by the European Community under the FP7 programme HadronPhysics3.
We acknowledge the support and infrastructure provided by the Trinity Centre for High Performance
Computing and the IITAC project funded by the HEA under the Program for Research in Third Level Institutes (PRTLI) co-funded by the Irish Government and the European Union.  
The work of CA and GA is carried out as part of the UKQCD collaboration and the STFC funded DiRAC Facility.
GA and CA are supported by STFC.  
GA is supported by the Royal Society, the Wolfson Foundation and the Leverhulme Trust. 
SK is grateful to STFC for a Visiting Researcher Grant and supported by the National Research
Foundation of Korea grant funded by the Korea government (MEST) No.\
2012R1A1A2A04668255.  
SR is supported by the Research Executive Agency (REA) of the European Union under Grant Agreement number PITN-GA-2009-238353 (ITN STRONGnet) and the Science Foundation Ireland, grant no.\ 11-RFP.1-PHY-3201.

\end{document}